\documentstyle[prl,aps,twocolumn]{revtex}

\input epsfig.sty

\begin{document}

\advance\textheight by 0.2in
\twocolumn[\hsize\textwidth\columnwidth\hsize\csname@twocolumnfalse\endcsname

\draft

\begin{flushright}
{\tt to appear in Phys. Rev. Lett.}
\end{flushright}

\title{Transverse thermal depinning and nonlinear sliding friction
of an adsorbed monolayer }
\author{Enzo Granato}
\address{Laborat\'orio Associado de Sensores e Materiais,
Instituto Nacional de Pesquisas Espaciais, \\
12201-190 S\~{a}o Jos\'e dos Campos, S\~ao Paulo, Brazil}
\author{S.C. Ying}
\address{Department of Physics,
Brown University, \\
Providence, Rhode Island 02912}

\maketitle

\begin{abstract}
We study the response of an adsorbed monolayer under a driving
force as a model of sliding friction phenomena between two
crystalline surfaces with a boundary lubrication layer. Using
Langevin-dynamics simulation,  we determine the nonlinear
response  in the direction transverse to a   high symmetry
direction along which the layer is already sliding. We find
 that below a finite transition
temperature, there exist a critical depinning force and
hysteresis effects in the transverse response in the dynamical
state when the adlayer is sliding smoothly along  the
longitudinal direction.
\end{abstract}

\pacs{68.35.Gy, 68.35.Rh, 81.40.Pq}

]

Driven lattice systems interacting with pinning potentials have
attracted growing interest recently due to the variety of
possible nonequilibrium dynamic phases. This problem is relevant
for many systems including adsorbed layers in tribology
\cite{exp,exp2,book,proc,persson,robbins,braun,gy}, charge-density
waves \cite{gruner} and  vortex lattices \cite{nori,vl,gled} in
type-II superconductors. In particular, in the boundary
lubrication problem for interfaces \cite{book}, the macroscopic
sliding friction is controlled by the response of the tightly
bound lubricant layer between the surfaces to the external driving
force. It has been shown that there are many dynamical phases,
and the transitions between these phases can lead to such
macroscopic behavior as the stick and slip motion. The essence of
the physics in the boundary lubrication has been reasonably well
understood with the study of a zeroth order model consisting in
an overlayer of interacting atoms adsorbed in a periodic
potential subject to an external driving force \cite{persson,gy}.
However, an interesting aspect of sliding friction which has not
been addressed so far is  how the nonlinear response of the
overlayer is affected by the symmetry properties of the
underlying periodic pinning. Initially, the overlayer can be
driven to slide along a high symmetry direction of the surface.
In this sliding state, the response to an additional force in the
transverse direction  is of interest. It  should provide
important information on the nature of the sliding state. In
addition, the transverse force can also generate novel
nonequilibrium dynamic phases of the overlayer. In this paper, we
report the result from a study of the transverse sliding friction
of a monolayer initially driven along the high symmetry direction
of the periodic pinning potential. We find evidence of a
transverse  depinning transition at a nonzero 
temperature $T_p$. The nonequilibrium dynamic phase below $T_p$
 shows  a critical depinning force  in the transverse
response even when the overlayer is in a state sliding smoothly
along the longitudinal direction.

We consider a  model of interacting adatoms  in a periodic
potential subject to an external driving force \cite{book,persson}.
The dynamics is described by the Langevin equation
\begin{equation}
m\ddot  {\bf r}_i + m \eta \dot{\bf r}_ i = - \frac{\partial
U}{\partial{\bf r}_i} - \frac{\partial V}{\partial{\bf r}_i} +
{\bf f}_i + {\bf F}
\end{equation}
where ${\bf r_i}$ is the adsorbate position, $U=\sum_i u(r_i)$ is
the periodic substrate pinning potential, $V=\sum_{i\ne j}
v(|{\bf r_i}-{\bf r_j})$ is the  interacting potential between
adatoms, ${\bf F}$ is the uniform external force acting on each
adatom and ${\bf f}_i$ is an uncorrelated  stochastic force, with
zero average, and variance related to the microscopic friction parameter $\eta$,
the mass of the particles $m$ and the temperature $T$ by the fluctuation
dissipation relation 
$<f_i^{\alpha}(t) f_i^{\alpha}(t')  > = 2 \eta m kT \delta(t-t')$,
where $\alpha$ represents the vector components. We
choose  a periodic pinning potential  with square symmetry
\begin{equation}
u({\bf r}) = U_o[2-\cos(2\pi x/a) - \cos(2 \pi y/a)]
\end{equation}
where $a$ is the lattice constant of the substrate and an
interacting Lennard-Jones pair potential
\begin{equation}
v(r_{ij}) = \epsilon [(r_o/r_{ij})^{12} - 2 (r_o/r_{ij})^6 ]
\end{equation}
where $\epsilon$ is the well depth and $r_0$ is the particle
separation at the minima in the pair potential. Periodic boundary
conditions are used, with $r_o/a=1.56$ and the adsorbate coverage
set to $\theta=1/2$. With this choice, the ground state of the
overlayer is a commensurate pinned $c(2 \times 2)$ structure which
has been extensively studied in connection with the  boundary
lubrication \cite{book,persson} problem and its longitudinal
response to an applied force is well understood. Thus, this pinned
structure provides  a convenient starting point for the present
study of transverse response. We vary the temperature $T$ and the
ratio $\epsilon/U_o$ that is essentially a measure of the
stiffness of the overlayer relative to the pinning potential.
Dimensionless units are used where $a=1$, $m=1$, $U_o=1$ and we
normalize the force by $2\pi U_o$  and velocities by $2\pi
U_o/\eta$. We study the nonlinear response of the overlayer
through dynamical simulations, using standard methods of Brownian
molecular dynamics \cite{allen}. As in the previous work on the
longitudinal response \cite{persson}, small systems are used in
order to allow for long equilibration times and reduce
statistical errors. A system consisting of $L \times L$ 
substrate atoms with $L= 10$ to $20$ was considered with the time variable
discretized in units of $\delta t=0.002 - 0.01 \tau$, where
$\tau=(ma^2/U_o)^{1/2}$. Calculations were performed  with
typically $ 1-4 \times 10^6$ time steps for each calculation of
time-averaged quantities allowing equal time steps for
equilibration.

As shown previously for an overlayer with the Lennard-Jones
interacting potential \cite{persson} and  also for a model with
pure elastic interacting potential \cite{gy}, the behavior of the
drift velocity ${\bf V_d}$ as a function of ${\bf F}$  along the
high symmetry direction $x$ or $y$ depends strongly on the
initial equilibrium phase at ${\bf F=0}$. We first summarize the
known behavior for the longitudinal response along the high
symmetry $x$-direction. When the overlayer at $ F=0 $ is in an
initial fluid or incommensurate state, then $V_d$ will be nonzero
for arbitrarily small external force leading to a finite sliding
friction $1/\bar \eta_L =V_d/F$ for arbitrarily small values of F.
For our choice of parameters, the initial state has a commensurate
$c(2\times2)$ structure below the melting  temperature $T_m$. For
this initial commensurate state, the drift velocity is
essentially zero below a critical value of F, neglecting the
contribution from creep motion due to thermal activation. The $F
\times V_d$ characteristics shows hysteresis behavior with
unequal critical forces $F_a$ and $F_b$ for increasing or
decreasing external force $F_x$, corresponding to the static and
kinetic friction forces,  respectively. This behavior is shown in
Fig. 1 for a temperature $T=0.2$ much below the melting
transition temperature $T_m=1.2$ of the commensurate overlayer at
$F=0$. In addition, at $F_b$ there is a velocity gap $V_b$. For
sliding velocities $V < V_b$, smooth sliding is not possible. As
a consequence, the motion of the overlayer shows stick and slip
behavior when driven by a spring moving with a velocity $V <
V_b$, in agreement with experimental observations where smooth
sliding is only found above a critical sliding velocity
\cite{exp}. 
As first pointed out by Persson \cite{persson}, for a sufficiently
strong interacting system, the particle velocity distribution 
relative to the center of mass is  Gaussian and one can define an 
effective temperature of the  overlayer $T^*$ ($kT^*=m(<v^2>-<v>^2)/2$). 
The irregular motion of the adatoms in the region $F_b < F_x < F_c$ 
leads to $T^*>T$, with the extra thermal dissipation, proportional 
to $T^*-T$, being needed to balance the heating effect of the external driving 
force to reach a steady state. The energy transfer to the overlayer is 
possible because the continuous translational symmetry is broken by the
external pinning potential. In the limit $F \rightarrow \infty $, the
pinning potential is negligible, and the overlayer slides rigidly 
with $T^*=T$. When the force is decreased, the sliding overlayer
undergoes a dynamic melting transition into a sliding liquid phase 
at $F_c$, where $T^*$ reaches a value above the melting transition 
temperature $T_m$. With further decrease
of the external force, the overlayer is finally pinned at the
critical kinetic frictional force $F_b$ to become  the commensurate
phase again.

In the sliding solid phase that exists at $F_x>F_c$ in Fig. 1, the
motion of the overlayer essentially averages out the pinning
potential. Therefore, for our choice of isotropic interacting
Lennard-Jones potential, the overlayer should have a triangular
structure although the pinning potential has a square symmetry.
However, in the transverse direction the pinning potential remains
essentially static in a reference frame comoving with the
overlayer \cite{gled} and it is plausible that the overlayer may
remain pinned in the transverse direction, depending on
temperature and velocity, although it is sliding in an
incommensurate state along the longitudinal $x$-direction.

We have determined the nonlinear response of the overlayer in the
transverse direction in the sliding state. Once the system has
reached a steady state for a fixed $F_x > F_c$, an additional
force  $F_y$ is applied  and  the average drift velocity 
${\bf V_d}$ is determined after the new steady state is reached. 
Alternatively, $\nu_t$ can also be obtained by applying a 
total force $(F_x, F_y)$ to an initial equilibrium state 
($F_x=0$). The two methods give the same result indicating a
well defined transverse depinning behavior \cite{gled} as 
shown in Fig. 2a. At temperatures below a critical value $T_p$, which
is much smaller than the melting transition temperature $T_m$,
the mobility $ \mu_{t}$ vanishes implying that the overlayer
remains  pinned along the transverse direction, even when  it is
sliding  along the longitudinal $x$-direction. Above $T_p$, the
mobility $ \mu_{t}$ has a finite value. In the temperature range
$T_p < T < T_m$, there is no transverse critical force and the
overlayer is an incommensurate sliding solid in all directions.
As shown in Fig. 2b and 2c, below $T_p$ the $F_y \times V_y$
characteristics shows a critical transverse force $F_{ya}$ and
hysteresis behavior when the system is first equilibrated at
lower temperatures. Note, however, that for decreasing transverse
force, the pinned state is only reached at $F_{yb}\sim 0$, unlike
the behavior in the longitudinal direction in Fig. 1, suggesting
that along the transverse direction the existence of a critical
force $F_{ya}$ may not necessarily lead to stick-and-slip motion.
The behavior of the nonlinear response in the transverse
direction correlates with the structure of the sliding state at
$F_y=0$. Fig. 3a,b and 3c,d show the instantaneous configurations
and the trajectories of the adatoms in  the overlayer below and
above $T_p $ for $F_y=0$. The trajectories are obtained by
superposing successive configurations in the sliding state for a
fixed time interval. It can be seen from Fig. 3c and 3d that in
the transverse pinned phase below $T_p$, the particles are
essentially moving in one-dimensional channels along the
principal $x$-axis of the pinning potential and the particle
coordination number is $4$, leading to a square lattice
structure. For $T>T_p$, the particle trajectories are free to
fluctuate  in the transverse direction (Fig. 3b) and an isotropic
triangular lattice structure occurs as shown in Fig. 3a. This is
exactly the structure that the overlayer would have in the
absence of a periodic pinning potential. The isotropic triangular
structure has also been observed in previous calculations
\cite{persson} but was considered the only possible phase in the
sliding state at any finite temperature. Here we find instead
that for temperatures below $T_p$ a transverse pinned phase with
different structure is the stable configuration. The dynamical
phase transition separating the two sliding states has the
characteristic features of an equilibrium first order transition. 
Besides the hysteresis effects mentioned earlier, we also observe 
supercooling effects in the sense that when the sliding 
state above $T_p$ is cooled down to a temperature below $T_p$,
at fixed $F_x$, the final sliding state retains its 
triangular structure in a metastable state as shown in Fig. 4a.
This indicates that a nucleation process with large energy barrier 
is involved. While the details change, the hysteresis and 
supercooling effects are observed up to the largest size 
($20 \times 20$) studied here. We also note that the physics
underlying the transverse depinning transition is very similar
to that of the longitudinal depinning transition which
has been shown to be very general and model independent 
\cite{persson,gy}. 

The results for other values of $F_x$ and
$\epsilon$ are qualitatively the same as long as $F_x > F_c$. The
linear transverse mobility $\mu_t$ ( in the limit $F_y
\rightarrow 0$ ) as a function of the longitudinal driving force
is shown in Fig. 4b for a temperature $T=0.2$  below $T_p$. It is
seen that $\mu_t$ remains zero for large $F_x$ and the onset of
finite $\mu_t$ for decreasing force occurs at $F_x=F_c$, where
the sliding state goes from a solid phase into a fluid phase as
would be expected from the behavior of the longitudinal response
in Fig. 1. So there is a transverse depinning transition at all
values of $F_x > F_c$. The dependence of the transverse depinning
transition temperature  $T_p$ on the stiffness of the overlayer
$\epsilon$ together with the melting transition temperature $T_m$
at $F_x=0$ are indicated in Fig. 4c. These transition
temperatures are located from the onset of finite linear mobility
$\mu_t$ and $\mu_L$ respectively for increasing temperatures.

Recent work \cite{nori} on the dynamics of sliding two dimensional
vortex lattices on a periodic potential also shows evidence of a
transverse pinning and critical Lorentz force  in the sliding
state. The form of interaction and pinning potentials for the
vortex lattice model are different from that for the overlayer
model considered here,  but this is not expected to be
qualitatively important in the sliding state \cite{lj}. More importantly,
the results for the vortex lattice were obtained at zero
temperature and for an overdamped dynamics with zero vortex mass.
Our results are obtained  for a nonzero mass, finite damping
coefficient $\eta$, and  finite temperatures. We have  shown that
in our model,  the transverse pinned state is stable against
thermal fluctuations up to a transverse depinning critical
temperature $T_p <  T_m $ as indicated in Fig. 4c. In view of the
similarities of these systems and the current interest on the
transverse critical current for vortex lattices in
superconductors \cite{gled}, it should be also interesting to
investigate how a nonzero mass \cite{mass} will affect the sliding state in
vortex lattices and if the presence of weak disorder will broaden
the sharp transition found in the present study.

The existence of a transverse static friction in the sliding state
implies that once two lubricated surfaces are sliding relative to
each other  in a particular direction commensurate with the
underlying pinning potential, changing the direction of motion
may require a nonzero additional force depending on the
temperature and sliding velocity. Experimentally, the transverse
critical force could in principle be detected from measurements of sliding
friction with simultaneous transverse (in plane) vibrations
between two boundary lubricated sliding surfaces, using similar
techniques as in  recent experiments which introduces  coupling
to normal (out-of-plane) vibrations \cite{exp2} and also in
sliding colloidal layers confined between two surfaces
\cite{ling}. Below the thermal depinning  temperature the
transverse vibrations should have very little effect on the
longitudinal motion for sufficiently small vibration amplitude,
while above the transition induced oscillations  on the
longitudinal friction are expected. We hope that this work will
motivate novel experiments on sliding friction. 

\bigskip
We thank X.S. Ling and B.N.J. Persson for helpful discussions. 
This work was supported  by a joint NSF-CNPq
grant and  FAPESP(99/02532-0)(E.G.).

\begin{figure}
\centering\epsfig{file=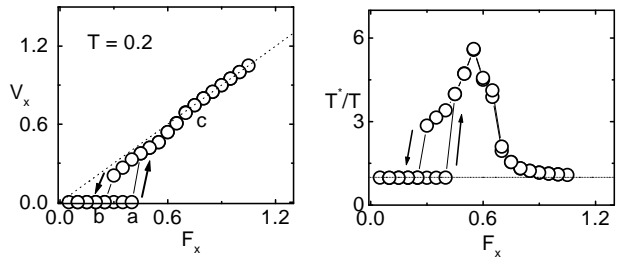,bbllx=1cm,bblly=12cm,bburx=20cm,
 bbury=26cm,width=8.5cm}
 
\caption{Drift velocity $V_x$ and effective temperature of the
overlayer $T^*$  as a function of the external force $F_x$  along
the $x$-axis for $\epsilon=1$ and $\eta=1$.
The direction of variation of $F_x$ is indicated by arrows. }
\end{figure}

\begin{figure}
\centering\epsfig{file=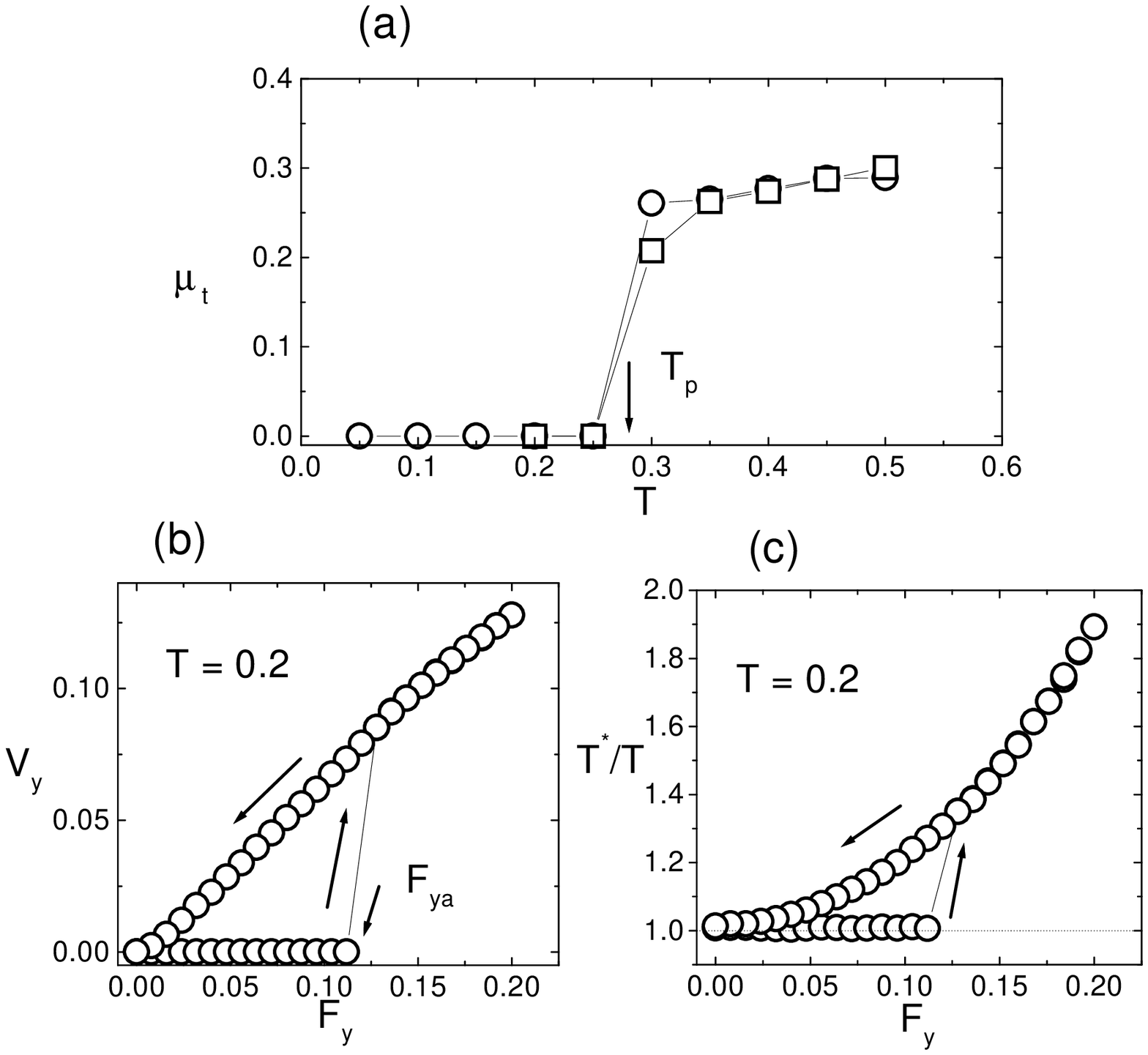,bbllx=1cm,bblly=5cm,bburx=20cm,
 bbury=26cm,width=8.5cm}
 
\caption{(a) Transverse mobility $\mu_t$ as a function of
temperature obtained by applying a total force ($F_x, F_y$) to an
initial equilibrium state (squares) and by applying and additional
$F_y$ to the sliding state at $(F_x, 0)$ (circles).
(b) and (c) Transverse velocity  $V_y$ and effective temperature
of the overlayer $T^*$ as a function of an
additional force along the $y$-axis for fixed $F_x$. 
For large $F_y$, the behavior of $T^*$ is smilar to Fig. 1. Large
arrows indicate the direction of force variation. Results are  
for $L=10$, $F_x=1.5$ and $\epsilon =1$. }
\end{figure}

\begin{figure}
\centering\epsfig{file=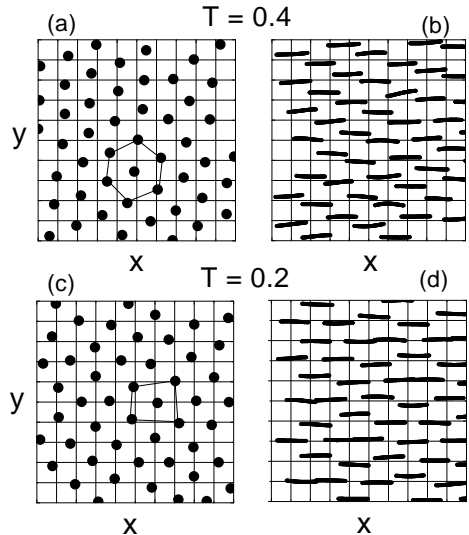,bbllx=1cm,bblly=8cm,bburx=20cm,
 bbury=26cm,width=8.5cm}

\caption{ Snapshot pictures  ((a) and (b) ) and trajectories  of
the adsorbates ((b) and (d)) at $F_y=0$ for temperatures $T=0.4$
and $T=0.2$  above and below the depinning temperature $T_p$,
respectively, for $F_x=1.5$ and $\epsilon=1$. Grid lines represent
the periodic pinning potential. }
\end{figure}

\begin{figure}
\centering\epsfig{file=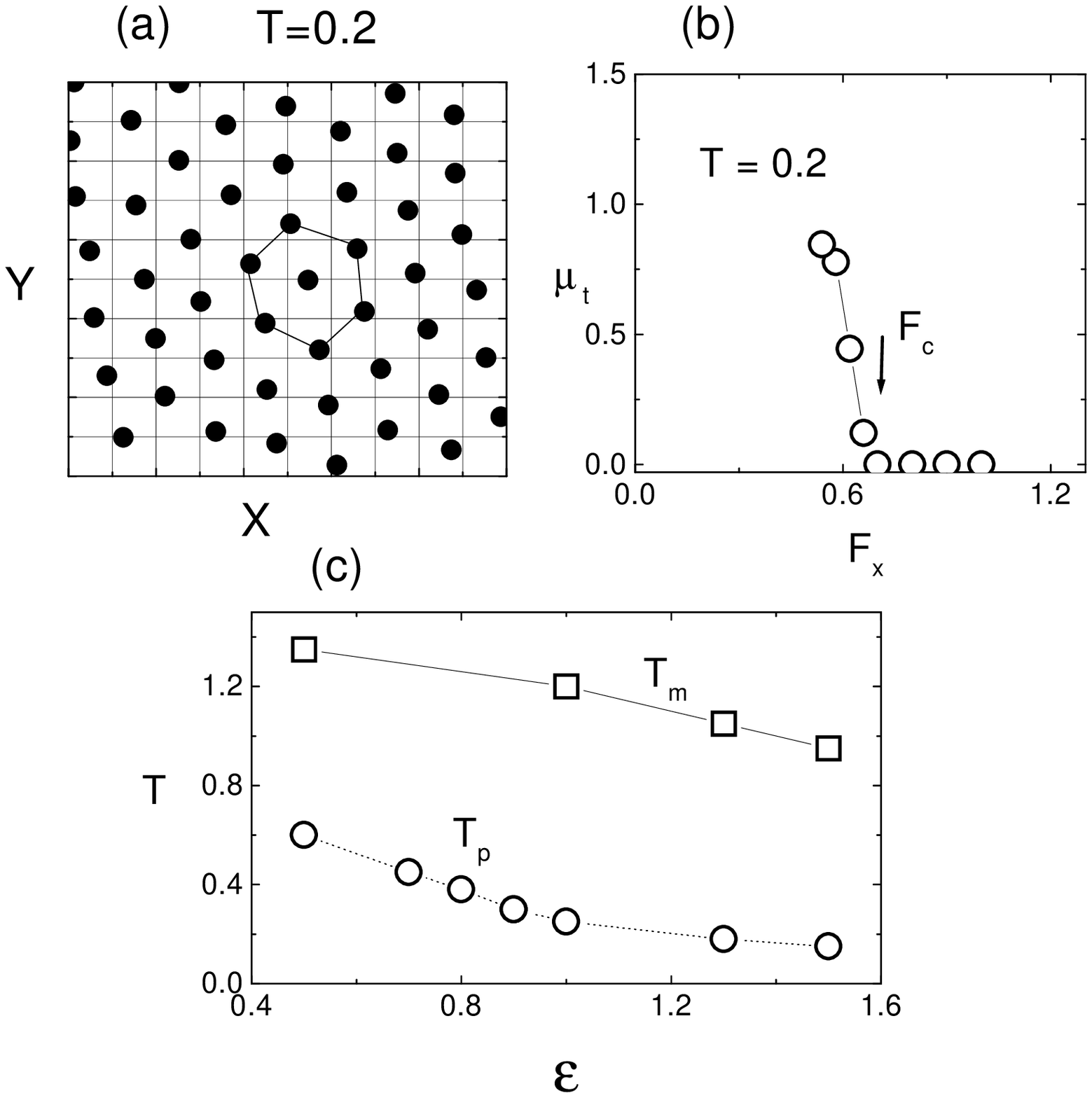,bbllx=1cm,bblly=5cm,bburx=20cm,
 bbury=26cm,width=8.5cm}

\caption{(a) Snapshot picture of the final state at $T=0.2$, below $T_p$,
obtained from an initial sliding state at $T=0.4$, above $T_p$, by lowering
the temperature at fixed $F_x=1.5$.
(b) Transverse mobility $\mu_t$ as a function of the
longitudinal force $F_x$ at a temperature  below the depinning
transition temperature $T_p$. Results are for $\epsilon =1$, $L=10$.
(c) Critical temperature $T_p$ for transverse depinning in
the sliding state for $F_x=1.5$ and  equilibrium ($F_x=0$)
melting transition temperature $T_m$ as a function  of the
overlayer stiffness $\epsilon$. }
\end{figure}

\begin{references}

\bibitem{exp} J.N. Israelachvili, Surf. Sci. Rep. {\bf 14}, 109 (1992);
H. Yoshizawa, P. McGuiggan, and J. Israelachvili, Science {\bf
259}, 1305 (1993); B. Bhushan, J.M. Israelachvili, and U.
Landaman, Nature (London) {\bf 374}, 607 (1995).

\bibitem{exp2} M. Heuberger, C. Drummond, and J. Israelachvili,
J. Phys. Chem. B {\bf 102}, 5038 (1998).

\bibitem{book} B.N. J. Persson, {\it Sliding Friction: Physical
Principles and Applications} (Springer, Heidelber, 1998).

\bibitem{proc} {\it Physics of Sliding Friction}, edited by B.N.J.
Persson and E. Tosatti (Kluwer, Dorbrecht, 1996).

\bibitem{persson} B.N.J. Persson, Phys. Rev. Lett. {\bf 71}, 1212
(1993); Phys. Rev. B {\bf 48}, 18140 (1993); J. Chem. Phys. {\bf
103}, 3449 (1995).

\bibitem{robbins} P.A. Thompson and M.O. Robbins, Science {\bf 250}, 792 (1990).

\bibitem{braun} O.M. Braun {\it et al.}, Phys. Rev. Lett. {\bf 78}, 1295 (1997).

\bibitem{gy} E. Granato and S.C. Ying, Phys. Rev. B {\bf 59}, 5154
(1999); E. Granato, M.R. Baldan, and S.C. Ying, in ref. \onlinecite{proc}, p. 103.

\bibitem{gruner}G. Gr\"uner, Rev. Mod. Phys. {\bf 60}, 1129 (1988).

\bibitem{nori} C. Reichhardt and F. Nori, Phys. Rev. Lett. {\bf 82},
414 (1999).

\bibitem{vl} J.Y. Lin {\it et al.}, Phys. Rev. B {\bf 54}, R12717
(1996)

\bibitem{gled} T. Giamarchi and P. Le Doussal, Phys. Rev. Lett. {\bf 76},
3408 (1996); P. Le Doussal and T. Giamarchi, Phys. Rev. B {\bf
57}, 11356 (1998).

\bibitem{allen} M.P. Allen and D.J. Tildesley, {\it Computer Simulation
of liquids} (Oxford University Press, New York, 1993).

\bibitem{lj} In contrast to the vortex potential, our pair potential
contains an attractive interaction for $r_{ij} > r_o$. However, additional
calculations using a modified potential consisting only of the repulsive part
of the Lennard-Jones potential give the same qualitatively results.

\bibitem{mass} N.B. Kopnin and V.M. Vinokur, Phys. Rev. Lett.
{\bf 81}, 3952 (1998).

\bibitem{ling} X.S. Ling, private communication.

\end{references}
\end{document}